\documentclass[aps,pre,twocolumn,groupedaddress,floatfix]{revtex4}
\usepackage{graphicx}
\usepackage{latexsym}
\usepackage{amsmath}
\usepackage{amsfonts}
\usepackage{epstopdf}
\usepackage{amssymb}
\usepackage{bm}
\usepackage{subfigure}
\usepackage{color} 
\usepackage{dcolumn}

\definecolor{red}{rgb}{1,0,0}

\renewcommand{\vec}{\bm}

\newcommand{\wi}{\mathrm{Wi}}

\graphicspath{{./plots/}}

\begin{document}


\title{Semidilute polymer solutions at equilibrium and under shear flow }

\author{Chien-Cheng Huang$^1$}
\author{Roland G. Winkler$^1$}
\email{r.winkler@fz-juelich.de} \affiliation{}
\author{Godehard Sutmann$^2$}
\author{Gerhard Gompper$^1$}
\affiliation{%
$^1$Theoretical Soft Matter and Biophysics, Institut f\"{u}r
Festk\"{o}rperforschung and Institute for Advanced Simulation,
Forschungszentrum J\"{u}lich, 52425 J\"{u}lich, Germany \\
$^2$J\"{u}lich Supercomputing Centre, Institute for Advanced
Simulation, Forschungszentrum J\"{u}lich, 52425 J\"{u}lich, Germany}%

\date{\today}

\begin{abstract}
The properties of semidilute polymer solutions are investigated at
equilibrium and under shear flow by mesoscale simulations, which
combine molecular dynamics simulations and the multiparticle
collision dynamics approach. In semidilute solution,
intermolecular hydrodynamic and excluded volume interactions
become increasingly important due to the presence of polymer
overlap. At equilibrium, the dependence of the radius of gyration,
the structure factor, and the zero-shear viscosity on the polymer
concentration is determined and found to be in good agreement with
scaling predictions. In shear flow, the polymer alignment and
deformation is calculated as function of concentration. Shear
thinning, which is related to flow alignment and finite polymer
extensibility, is characterized by the shear viscosity and the
normal stress coefficients.
\end{abstract}


\maketitle


\section{Introduction}
The properties of dilute polymer solutions under shear flow have
been studied
intensively~\cite{smith:99,schr:05,teix:05,gera:06,doyl:00,cela:05,cher:05,puli:05,schr:05_1,delg:06,wink:06_1,wink:04,ripo:06,aust:99,wang:90}.
Recent advances in experimental single-molecule techniques even
provide insight into the dynamics of individual polymers under
shear
flow~\cite{smith:99,schr:05,teix:05,schr:05_1,gera:06,doyl:00}.
Similarly, the dynamics of individual polymers in a melt has been
addressed
extensively~\cite{mcle:02,kroe:04,rubi:03,kim:09,jose:07,jose:08}.
However, we are far from a similar understanding of the dynamics
of semidilute polymer solutions, although insight into the
behavior of such systems is of fundamental importance in a wide
spectrum of systems ranging from biological cells, where transport
appears in dense environments \cite{elli:03}, to turbulent drag
reduction in fluid flow. Moreover, in semidilute solutions of long
polymers, viscoelastic effects play an important role. Due to the
long structural relaxation time, the internal degrees of freedom
of a polymer cannot relax sufficiently fast under non-equilibrium
conditions and an elastic restoring force tries to push the system
towards its original state. Here, a deeper understanding can be
achieved by mesoscale hydrodynamic simulations
\cite{kapr:08,gomp:09}.

The dynamical behavior of dilute and semidilute polymer solutions
is strongly affected or even dominated by hydrodynamic
interactions~\cite{doi:86,kapr:08,gomp:09}. From a theoretical
point of view, scaling relations predicted by the Zimm model at
infinite dilution, e.g., for the dependence of dynamical
quantities as viscosity and relaxation time on the polymer length,
are, in general, accepted and
confirmed~\cite{doi:86,ahlr:99,muss:05}. However, as the
concentration of the polymer is increased beyond the segmental
overlap concentration $c^*$, where the volume occupied by polymer
coils is equal to the total volume, the dynamics becomes more
complex due to intermolecular excluded volume
interactions~\cite{dege:79,doi:86,rasp:95,pate:92}. For this
regime, the scaling theory established by de Gennes describes the
polymer dynamics using the concept of 'blobs'~\cite{dege:79}. Here
a blob consists of $g$ monomers and has the radius $\xi$. A
polymer chain comprised of $N_m$ monomers can be regarded as
composed of $N_m/g$ blobs which are hydrodynamically independent.
Inside of a single blob, the dynamics follows the predictions of
the Zimm model in dilute solution. On length scales larger than
$\xi$, hydrodynamic and excluded volume interactions are screened
due to chain overlap. Thus, the polymer dynamics on this scale can
be described by the Rouse model. When the concentration is further
increased, the polymer dynamics is dominated by entanglement
effects, which arise from physical uncrossability of chain
segments for sufficiently long polymers. Based on this theory, the
relaxation time and also the zero-shear viscosity  in the
semidilute regime can be scaled by using the concentration ratio
$c/c^*$, where $c$ is the segment concentration. Various
experiments confirm the predicted dependencies for the relaxation
time and viscosity~\cite{pate:92,taka:85,rasp:95,adam:83,heo:05}.
However, a systematic simulation study is still lacking, even
though single-chain hydrodynamic simulations are well
established~\cite{pier:91,dunw:91,pier:92,ahlr:99,lyul:99,pete:99,jend:02,hsie:04,muss:05,ryde:06,send:07,zhan:09}.
The difficulty is that in the semidilute regime a large polymer
overlap is necessary, whereas at the same time the segmental
density has to be rather low to retain hydrodynamic interactions,
which requires the simulation of long polymers \cite{ahlr:01}.


The large length- and time-scale gap between the solvent and
macromolecular degrees of freedom requires a mesoscale simulation
approach in order to assess their structural, dynamical, and
rheological properties. Here, we apply a hybrid simulation
approach, combining molecular dynamics simulations (MD) for the
polymers with the multiparticle collision dynamics (MPC) method
describing the
solvent~\cite{male:99,male:00,male:00_1,kapr:08,ripo:04,ripo:05,muss:05,gomp:09,padd:06}.
As has been shown, the MPC method is very well suited to study the
non-equilibrium properties of polymers
\cite{wink:04,ryde:06,cann:08,chel:10}, colloids
\cite{ripo:06,padd:04,wyso:09}, and other soft-matter object such
as vesicles \cite{nogu:04} and cells \cite{nogu:05,mcwh:09} in
flow fields.

Experiments~\cite{smith:99,ledu:99,schr:05,teix:05,schr:05_1,gera:06,doyl:00},
theoretical studies~\cite{wang:90,wink:06_1}, and
simulations~\cite{liu:89,pier:95,pete:99,aust:99,pier:00,hsie:04,stol:05,ryde:06}
of individual polymers under shear-flow conditions exhibit large
deformations and a strong alignment of the polymers. Moreover, a
large overlap is present in a semidilute solution of long
polymers. A typical simulation requires $10^5$ - $10^6$ monomers
and $10^7$ - $10^8$ fluid particles. Hence, despite the adopted
mesoscale approach, large systems can only be studied on a
massively parallel computer architecture. Here, we present results
of large-scale simulations of semidilute polymer solutions under
shear. The simulations were performed with our program MP$^2$C
(massively parallel multiparticle collision
dynamics)~\cite{sutm:10}, which exhibits excellent scaling
behavior on the massively parallel architecture of the IBM Blue
Gene/P computer \cite{sutm:11}. For the MPC fluid, we find a
linear increase of the speedup with increasing number of cores in
a strong scaling benchmark up to $2^{12}, \ 2^{14}, \ 2^{16}$
cores for $\ 10^7, \ 8 \times 10^7, 6 \times 10^8$ fluid
particles, respectively.

The paper is organized as follows. In Sec. II, the model and
simulation approache are described. The equilibrium properties of
the system are presented in Sec. III. Sec. IV is devoted to the
structural and conformational properties of the system under
stationary shear flow. In Sec. V, results are presented for the
rheological properties and finally, Sec. VI summarizes or
findings.


\section{Model and Parameters}

The solution consists of $N_p$ linear flexible polymer chains embedded in an
explicit solvent. A linear polymer is composed of $N_m$ beads of
mass $M$ each, which are connected by harmonic springs. The bond
potential is
\begin{equation}
U_b=\frac{\kappa}{2}\sum^{N_m-1}_{i=1}
\left( \vert \vec{r}_{i+1}-\vec{r}_i\vert - l\right)^2,
\end{equation}
where $l$ is the bond length and $\kappa$ the spring constant.
Inter- and intramolecular excluded-volume interactions are taken
into account by the repulsive, shifted, and truncated
Lennard-Jones potential
\begin{equation}
U_{LJ}= 4\epsilon\left[\left(\frac{\sigma}{r}\right)^{12}-\left(\frac{\sigma}{r}\right)^{6}+\frac{1}{4} \right]\Theta\left(2^{1/6}\sigma-r\right),
\end{equation}
where $\Theta(x)$ is the Heaviside function [$\Theta(x) =0$ for $x
<0$  and $\Theta(x) = 1$ for $x\geqslant0$]. The dynamics of the
chain monomers is determined by Newton's equations of motion, which are
integrated by the velocity Verlet algorithm with the time step
$h_p$ \cite{alle:87}.

The solvent is simulated by the multiparticle collision (MPC)
dynamics method \cite{male:99,male:00,kapr:08,gomp:09}. It is
composed of $N_s$ point-like particles of mass $m$, which interact
with each other by a stochastic process.  The algorithm consists
of alternating streaming and collision steps. In the streaming
step, the particles move ballistically and their positions are
updated according to
\begin{equation}
\vec{r}_i(t+h)=\vec{r}_i(t)+h\vec{v}_i(t) ,
\end{equation}
where $i =1, \ldots, N_s$ and $h$ is the time interval between
collisions. In the collision steps, the particles are sorted into
cubic cells of side length $a$ and their relative velocities, with
respect to the center-of-mass velocity of the cell, are rotated
around a randomly oriented axis by a fixed angle $\alpha$, i.e.,
\begin{equation}
\vec{v}_i(t+h)=\vec{v}_i(t)+(\mathbf{R}(\alpha)-\mathbf{E})(\vec{v}_i(t)-\vec{v}_{cm}(t)),
\end{equation}
where $\vec{v}_i(t)$ denotes the velocity of particle $i$ at time
$t$, $\mathbf{R}(\alpha)$ is the rotation matrix, $\mathbf{E}$ is
the unit matrix, and
\begin{equation}
\vec{v}_{cm} = \frac{1}{N_c}\sum^{N_c}_{j=1}\vec{v}_j,
\end{equation}
is the center-of-mass velocity of the particles contained in the
cell of particle $i$. $N_c$ is the total number of solvent
particles in that cell.

The solvent-polymer coupling is achieved by taking the monomers
into account in the collision step, i.e., for collision cells containing monomers, the center-of-mass velocity reads
\begin{equation}
\vec{v}_{cm}(t)=\frac{\sum^{N_c}_{i=1}m\vec{v}_i(t)+\sum^{N_c^m}_jM\vec{v}_j(t)}{mN_c+MN_c^m},
\end{equation}
where $N^m_c$ is the number of monomers within the considered
collision cell. To insure Galilean invariance, a random shift is
performed at every collision step~\cite{ihle:01}. The collision
step is a stochastic process, where mass, momentum and energy are
conserved, which leads to the build up of correlations between the
particles and gives rise to hydrodynamic interactions.

To impose a shear flow, for the short chains, we apply
Lees-Edwards boundary conditions \cite{alle:87}. A local
Maxwellian thermostat is used to maintain the temperature of the
the fluid at the desired value~\cite{huan:10}.

A parallel MPC algorithm is exploited for systems of long chains,
which is based on a three-dimensional domain-decomposition
approach, where particles are sorted onto processors according to
their spatial coordinates \cite{sutm:10}. Here, shear flow is
imposed by the opposite movement of two confining walls. The walls
are parallel to the $xy$-plane and periodic boundary conditions
are applied in the $x$- and $y$- directions. The equations of
motion of the solvent particles are modified by the wall
interaction~\cite{wink:09}. We impose no-slip boundary conditions
by the bounce-back rule, i.e., the velocity of a fluid particle is
reverted when it hits a wall and phantom particles in a wall are
taken into account. The same rule is applied for monomers when
colliding with a wall~\cite{gomp:09}.

The simulation parameters are listed in Table \ref{table1}. All
simulation are performed with the rotation angle
$\alpha=130^{\circ}$. Length and time are scaled according to
$\tilde{r}_{\beta}=r_{\beta}/a$, $\beta \in \{x,y,z\}$, and
$\tilde{t}=t\sqrt{k_BT/(ma^2)}$, which corresponds to the choice
$k_BT=1$, $m=1$, and $a=1$, where $T$ is the temperature and $k_B$
the Boltzmann constant. The collision time is $\tilde h=0.1$. The
parameters yield the shear viscosity $\tilde \eta = \eta/
\sqrt{mk_BT/a^4} = 8.7$. A large rotation angle $\alpha \gtrsim
90^{\circ}$ and a small time step $h$ are advantages to obtain
high fluid viscosities, low Reynolds numbers, and larger Schmidt
numbers. The selected values yield the fluid Schmidt number $Sc
\approx 14$, i.e., a fluid is simulated rather than a gas
\cite{ripo:05,gomp:09}. Between MPC collisions, the monomer
dynamics is determined by molecular dynamics simulations for
$h/h_p$ steps, with $\tilde{h}_p=0.002$. Moreover, we choose
$l=a$, $\sigma=a$, $k_BT/ \epsilon =1$, and $\tilde{\kappa}=\kappa
a^2 /(k_BT)=5\times10^3$. The large spring constant
$\tilde{\kappa}$ ensures that the mean of the bond length changes
by less than $0.5\%$ and the variance of the bond length
distribution by $3\%$ only, even at the largest shear rate.

\begin{table*}[t]
\caption{\label{table1} List of simulation parameters. $L_x$,
$L_y$, $L_z$ denote the dimensions of the simulation box, $\dot
\gamma$ the shear rate, and $\left\langle N_c \right\rangle$ is
the mean number of fluid particles in a collision cell. For $N_p$,
$c$, $c^*$, and $\dot \gamma$ the smallest and largest values used
are given. Actual concentrations are provided in figure captions.}
\begin{ruledtabular}
\begin{tabular}{ccccccccc}
$N_m$ & $N_p$ & $L_x/a$$\times$ $L_y/a$$\times$ $L_z/a$ &$\langle R_{g0}^2\rangle/l^2$& $c/l^{-3}$ & $c^*/l^{-3}$&$c/c^*$ & $\dot{\gamma}/\sqrt{k_BT/(ma^2)}$ &$\langle N_c\rangle$ \\
\hline
20& 10 -- 200& 20$\times$20$\times$20 &7.05& 0.025 -- 0.5 &0.26&0.098 -- 1.96 & $7.5\times10^{-4}$ &5\\
40& 10 -- 100 &  20$\times$20$\times$20 &17.29& 0.05 -- 0.5 &0.13&0.38 -- 3.76 & $7.5\times 10^{-4}$ &5 \\
50& 10 -- 512 &  50$\times$50$\times$50 &24.51& 0.004 --  0.205&0.098&0.041 -- 2.08 & $10^{-4} - 3 \times 10^{-1}$ &10 \\
250&50 -- 3000 &  450$\times$75$\times$75 &163.49&  0.0049 -- 0.296&0.029&0.17 -- 10.38 & $10^{-6} - 3\times 10^{-2}$&10\\
\end{tabular}
\end{ruledtabular}

\end{table*}

\section{Equilibrium properties}
\label{sec_equil}

Before we will address polymer solutions under shear, the scaling
behavior of equilibrium properties is discussed, in order to
determine the crossover at which a solution starts to follow the
expected scaling laws of a semidilute solution.

\subsection{Conformational properties}

The mean square radius of gyration $\langle R^2_{g0} \rangle$ in
dilute solution obeys the scaling relation
\begin{equation} \label{gyrat}
\langle R^2_{g0} \rangle  \propto N_m^{2\nu} ,
\end{equation}
with an exponent $\nu\approx$ 0.59 for a good solvent
\cite{dege:79,doi:86}. The obtained values for $\langle
R^2_{g0}\rangle$ are listed in Table~\ref{table1} for various
chain lengths. A fit of Eq.~(\ref{gyrat}) to our simulation data
obtained at the lowest concentrations (cf. Table 1) yields the
exponent $\nu$=0.61, in good agreement with theory and
experimental data~\cite{dege:79,doi:86}.

As the concentration increase, the polymer coils start to overlap
when the monomer concentration $c=N_m N_p/V$ exceeds the value
$c^* = N_m/V_p$, with the volume of a polymer $V_p = 4 \pi \langle
R_{g0}^2\rangle^{3/2}/3$  and the totally available volume $V$
\cite{doi:86}. Scaling considerations predict the dependence
\begin{equation} \label{eq:Rg_c}
\langle R_g^2 \rangle =\langle R_{g0}^2 \rangle \left(\frac{c}{c^*}\right)^{(2\nu-1)/(1-3\nu)}
\end{equation}
for the radius of gyration at concentrations $c \gg c^*$
\cite{dege:79,doi:86}. This relation has been confirmed
experimentally~\cite{daou:75} and by computer simulations
\cite{stol:05,peli:08}.

\begin{figure}[t]
\begin{center}
\includegraphics*[width=.4\textwidth,angle=0]{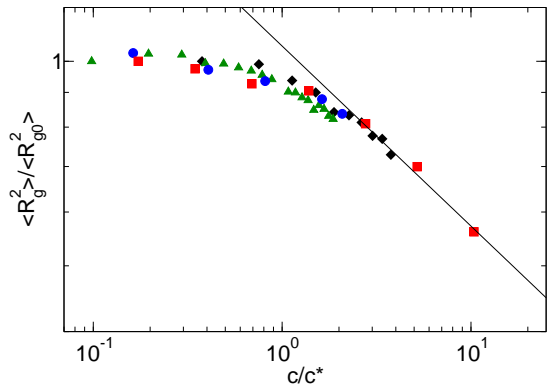}
\caption{Relative mean square radii of gyration as a function of
the scaled concentration $c/c^*$ for the chain lengths $N_m=20$ (${\color{green}
\blacktriangle }$), $40$ ($\blacklozenge$), $50$ (${\color{blue} \bullet }$),
and $250$  (${\color{red} \blacksquare }$).
The solid line indicates the dependence of Eq.~(\ref{eq:Rg_c}) for $\nu=0.61$. }
\label{Rg_Rg0}
\end{center}
\end{figure}

Figure \ref{Rg_Rg0} shows radii of gyration for various polymer
lengths and concentrations. Our simulation results follow the
scaling predictions. For $c\ll c^*$, $\langle R_g^2\rangle$ is
independent of polymer concentration. At $c/c^* \approx 1$, the
coil size starts to decrease and for $c \gg c^*$,  $\langle
R_g^2\rangle \sim c^{-0.265}$ with $\nu = 0.61$.

As suggested by de Gennes, the coil overlap implies a screening of
excluded volume and hydrodynamic interactions on length scales
larger than the blob size $\xi$ \cite{dege:79,doi:86}. Below this
length, the swollen conformations and hydrodynamic interactions
are maintained. The correlation length $\xi$ is independent of
chain length $N_m$ and is only a function of monomer concentration
at strong overlap, which yields the scaling relation
\begin{equation} \label{eq:xi_c}
\xi=\langle R_{g0}^2 \rangle^{1/2} \left(\frac{c}{c^*}\right)^{-\nu/(3\nu-1)}
.
\end{equation}

The crossover is reflected in the polymer structure factor
\begin{equation}
S(\vec{q}) = \frac{1}{N_m}\sum^{N_m}_{i,j=1}\langle
\exp[-i \vec{q} (\vec {r}_i-\vec{r}_j)]\rangle ,
\end{equation}
which exhibits the power-law dependence $S(\vec{q}) \sim
q^{1/\delta}$ for $1 \ll q \langle R_g^2 \rangle^{1/2} \ll \langle
R_g^2 \rangle^{1/2}/ l$ with $\delta=1/2$ for a melt and
$\delta=\nu$ in good solvent \cite{stro:07}. Hence, in a
semidilute solution two regimes are expected, separated by the
correlation length $\xi$: A good solvent behavior on length scales
smaller than $\xi$ and a Gaussian chain behavior on length scales
larger than $\xi$, i.e.,
\begin{equation}
S(q) \sim  \left\{ \begin{array}{ccc}
q^{-2} & \mbox{for}
& 2\pi/\langle R_g^2 \rangle^{1/2}   < q < 2\pi/ \xi, \\
q^{-1/\nu} &  \mbox{for} & 2\pi/ \xi < q < 2\pi / l,
\end{array}\right. .
\end{equation}

\begin{figure}[t]
\includegraphics*[width=.4\textwidth,angle=0]{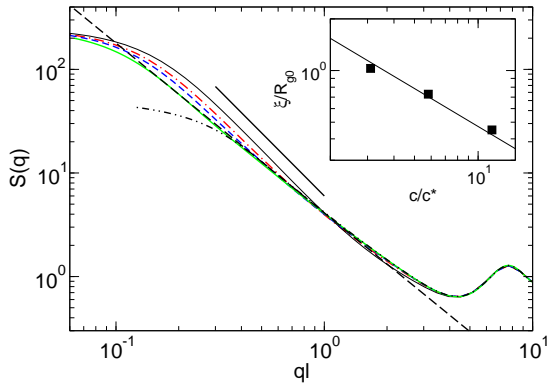}
\caption{\label{fig1} Structure factors of polymers of length
$N_m=250$ for the concentration ratios $c/c^*= 0.17$ ({\color{green}-----}),
$c/c^*= 2.77$
({\color{blue}- - -}), $c/c^*= 5.19$ ({\color{red}- $\cdot$ -}),
and $c/c^*= 10.38$ (-----), and
of $N_m=50$ in the dilute regime (- $\cdot$$\cdot$ -).
The dashed and solid straight lines represent power-law functions with the
exponents $1/\nu = 1/0.61$  and $2$, respectively. Inset: Dependence of the
blob size on the concentration. }
\end{figure}

In Fig.~\ref{fig1} polymer structure factors are shown  for $N_m=
250$ in dilute and semidilute solutions as well as for  $N_m =50$
in dilute solution. In order to obtain the two regimes separated
by $\xi$, the polymer chains have to be sufficiently long to
provide not only a ratio $c/c^*\gg 1$ but also a low segment
concentration $c$. As shown in the figure, for dilute solutions,
$S(q)$ decays with an exponent $1/\nu$, where $\nu =0.61$. The
polymers in the semidilute regime show a crossover from the
scaling behavior $S \sim q^{-2}$ at small $q$ to the behavior $S
\sim q^{-1/\nu}$ at large $q$ values. The crossover between the
two regimes corresponds to $q \approx 2\pi/\xi$. The values for
$\xi$ are presented in the inset of Fig.~\ref{fig1} and are found
to be in excellent agreement with the scaling
prediction~(\ref{eq:xi_c}) with $\nu = 0.61$. Thus, the scaling
relation captures the concentration dependence of the blob size
for the considered range very well.

\subsection{Dynamics}
\label{equ_dynamics}

The polymer dynamics is dominated by hydrodynamic interactions in
dilute solution. Theoretical results on the concentration
dependence of the relaxation times for small overlap
concentrations are presented in Refs.~\cite{muth:78,muth:84}.
Compared to the infinite-dilution limit, a term linear in the
concentration is obtained, which is consistent with experimental
data~\cite{pate:92}. However, the experimental data can also well
be fitted by an empirical exponential function~\cite{pate:92}.

In semidilute solution, the intermolecular interactions between
different chains become increasingly important. The dynamics of
the polymers can be classified according to their intermolecular
interactions as unentangled or entangled~\cite{pate:92,rasp:95}.
In the {\em unentangled regime}, the monomers move according to
Brownian motion in all three spacial directions and their dynamics
can be described by the Rouse behavior of polymers which  consist
of blobs. Thus, the longest relaxation relaxation time reads as
\begin{equation}
\tau=\tau_{b} \left(\frac{N_m}{g} \right)^2,
\label{t_blob_rouse}
\end{equation}
where $\tau_{b}$ is the blob relaxation time and $g$ the number of
monomers in a blob. Inside of a blob, the dynamics follows the
Zimm behavior
\begin{equation}
\tau_{b} \sim \left( \frac{\xi}{l} \right)^3 ,
\label{t_blob_zimm}
\end{equation}
and the longest relaxation time exhibits the concentration
dependence
\begin{equation}
\tau = \tau_0 \left(\frac{c}{c^*} \right)^{(2-3\nu)/(3\nu-1)} .
\label{t_unentangled}
\end{equation}

In the \textit{entangled regime}, polymers are assumed to exhibit reptation
motion inside a tube caused by the presence of neighboring chains.
The monomer dynamics is then described by reptation
theory~\cite{dege:79}, where the longest relaxation time obeys the
relation
\begin{equation}
\tau \sim \left(\frac{c}{c^*} \right)^{(3-3\nu)/(3\nu-1)}.
\label{t_entangled}
\end{equation}

\begin{figure}[t]
\begin{center}
\includegraphics*[width=.4\textwidth,angle=0]{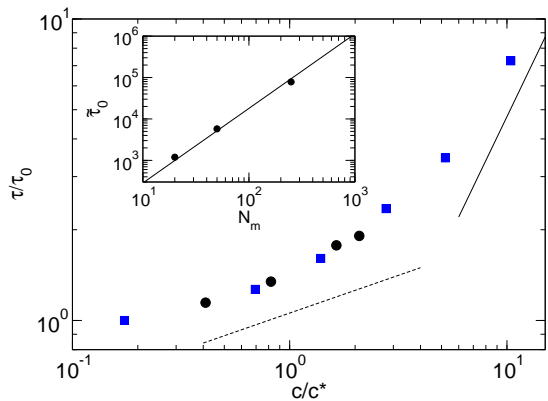}
\caption{Concentration dependence of longest polymer relaxation times $\tau$ for
the polymer lengths $N_m=50$ ($\bullet$) and $250$ (${\color{blue} \blacksquare }$).
The dashed line indicates the prediction
Eq.~(\ref{t_unentangled}) and the continues line Eq.~(\ref{t_entangled})
 with $\nu=0.6$. Inset: Polymer length dependence of the relaxation time at
 infinite dilution. The solid line shows the dependence
 $\tau_0 \sim N_m^{3 \nu}$ with $\nu = 0.6$.
\label{Relaxation_c}}
\end{center}
\end{figure}

By calculating end-to-end vector correlation functions, which
exhibit an exponential decay, we determined the longest polymer
relaxation times $\tau$ for various concentrations. The relaxation
time $\tau_0$ at infinite dilution  is obtained by extrapolation
to zero concentration. The obtained values for $\tau_0$ are shown
as a function of polymer length in the inset of
Fig.~\ref{Relaxation_c}. Their length dependence is well described
by the power-law $\tau_0 \sim N_m^{3 \nu}$, with $\nu=0.6$, in
accord with predictions of the Zimm model \cite{doi:86}.

Figure~\ref{Relaxation_c} depicts the dependence of the relaxation
time on concentration. In the vicinity of $c/c^* = 1$, $\tau$
follows the scaling prediction (\ref{t_unentangled}) for an
unentangled semidilute polymer solution. With increasing
concentration, $\tau$ increases faster, which we attribute to
strong intermolecular interactions. Although, there are no
entanglements in our system for $c/c^*<10$, the predicted
dependence for entangled polymer melts is indicated by the solid
line for illustration. This dependence is not reached and requires
longer polymers or higher concentrations. A very similar
dependence has been obtained experimentally in Ref. \cite{pate:92}
over comparable concentration and relaxation time ranges.

\begin{figure}[t]
\begin{center}
\includegraphics*[width=.4\textwidth,angle=0]{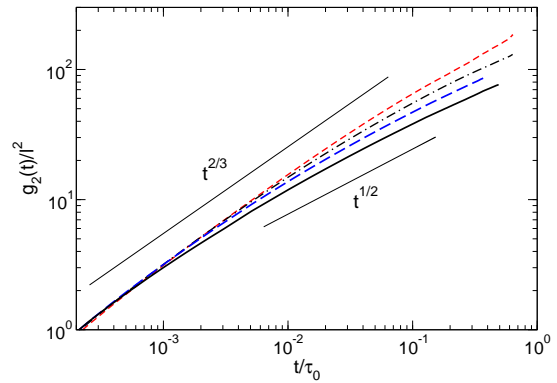}
\caption{Mean square displacements of monomers in the center-of-mass
reference frame for the concentrations $c/c^*=0.17$ ({\color{red}- - -}),
$c/c^*=2.77$ (- $\cdot$ -), $c/c^*=5.19$ ({\color{blue}-- -- --}),
$c/c^*=10.38$ (-----) of polymers of length $N_m=250$.
The short lines indicate the dependencies $g_2 \sim t^{2/3}$ and
$g_2 \sim t^{1/2}$, respectively.
\label{msd_9}}
\end{center}
\end{figure}

According to the Zimm model~\cite{zimm:56,doi:86,harn:96,muss:05},
hydrodynamic interactions strongly affect the diffusive
dynamics of polymers in solution and lead to the time dependence
\begin{equation}
g_2(t)=\langle ([{\bm r}_i(t)-{\bm r}_{cm}(t)]-[{\bm r}_i(0)-{\bm r}_{cm}(0)])^2 \rangle \sim t^{2/3}
\label{g2_dilute}
\end{equation}
of their monomer mean squared displacement  in the center-of-mass
reference frame for time scales larger than the Brownian time
\cite{dhon:96} and smaller than the longest relaxation time at
which $g_2(t)$ saturates. In the semidilute regime, hydrodynamic
interactions are screened for time scales larger than $\tau_b$,
the time needed to diffuse a blob diameter \cite{ahlr:01}.
Consequently,  after a time $\tau_b$, the dynamics is Rouse-like
and $g_2(t) \sim t^{1/2}$~\cite{ahlr:01}. Figure~\ref{msd_9}
displays $g_2(t)$ for various concentrations for polymers of
length $N_m=250$.

For a dilute solution with  $c/c^*=0.17$, $g_2(t)\sim t^{2/3}$ in
the time interval $10^{-3} <  t/\tau_0 < 10^{-1}$ as expected. For
$t > \tau$, $g_2$ slowly approaches a plateau value. At larger
concentrations, $g_2(t)$ displays a $t^{2/3}$ dependence which
turns into a $t^{1/2}$ behavior at a time $\tau_b$, which decrease
with increasing concentration. The concentration dependence of the
mean squared displacement reflects the screening of hydrodynamic
interactions in the semidilute regime. However, the dependence of
$\tau_b$ on concentration, which is linked to the screening length
$\xi_{H}$ according to $\tau_b \sim \xi^3_{H}$ and $\xi_{H} \sim
c^{-\gamma}$, where $\gamma$ is predicted to be $1$
\cite{free:74,edwa:84}, $0.6$ \cite{fred:90}, or $0.5$
\cite{dege:79,ahlr:01}, respectively, cannot be obtained
unambiguously from our simulations, because the different time
regimes are too short. Simulations of longer polymers are
necessary to arrive at clear and pronounced diffusion regimes.

\section{Semidilute polymer solutions in shear flow}





\begin{figure}
  \includegraphics*[clip,angle=270,width=0.4\textwidth]{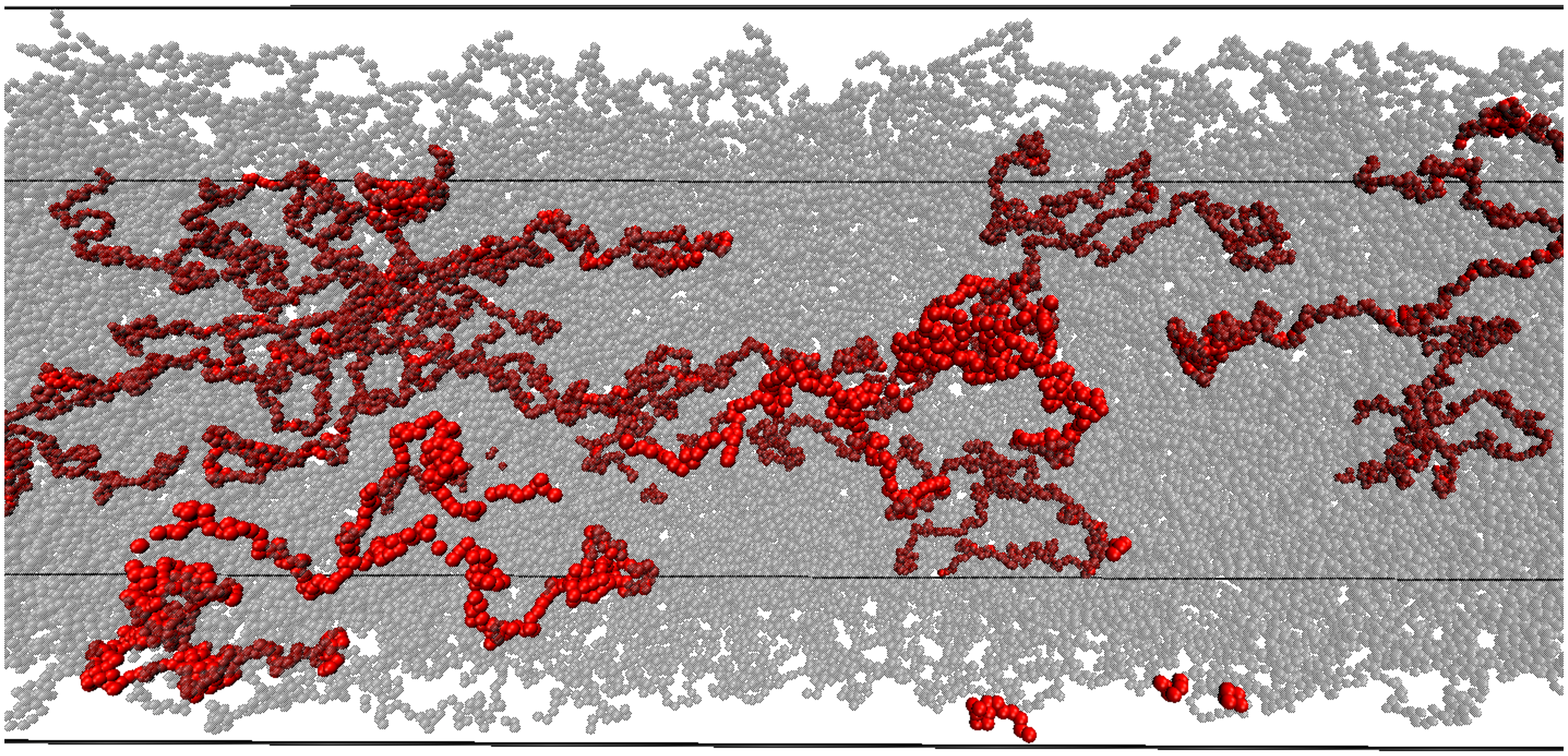}\\[1ex]
  \includegraphics*[clip,angle=270,width=0.4\textwidth]{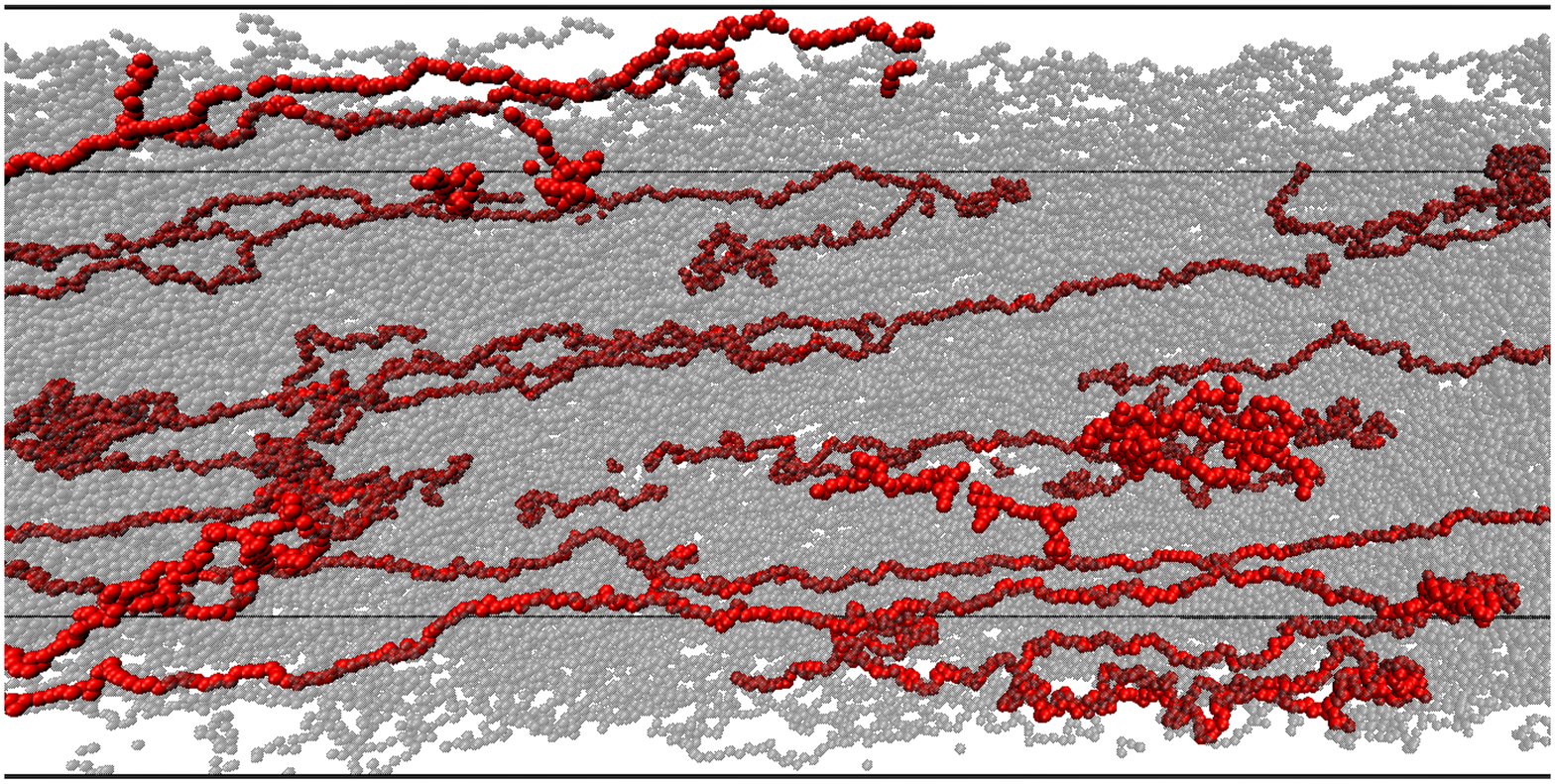}
  \caption{Snapshots of systems with $N_p=800$ polymers of length $N_m=250$
  for the Weissenberg
  numbers $\mathrm{Wi_c}=18$ (top) and $\mathrm{Wi_c}=184$ (bottom).
  For illustration, some of the chains are highlighted in red.
  \label{fig:snapshots}}
\end{figure}

We now discuss the properties of polymer solutions in shear flow.
At infinite dilution, the flow strength is characterized by the
Weissenberg number $\mathrm{Wi}=\dot{\gamma}\tau_{0}$, where
$\dot{\gamma}$ is the bare shear rate. For $\mathrm{Wi}\ll 1$, the
weak shear flow regime, the chains are able to undergo
conformational changes before the local strain has changed by a
detectable amount, while for $\mathrm{Wi}\gg 1$, the chains are
driven by the flow and they are not able to relax back to the
equilibrium conformation. This is illustrated in
figure~\ref{fig:snapshots}, which displays snapshots for various
flow rates. At small Weissenberg numbers, the polymers are only
weakly perturbed and are close to their equilibrium conformations,
whereas large $\rm Wi$ imply large deformations and a strong
alignment with flow.

As pointed out in section~\ref{equ_dynamics}, the polymer
relaxation times depend on concentration. Thus, in the following,
some properties will be characterized by the Weissenberg number
$\mathrm{Wi}_c= \mathrm{Wi} \, \tau(c)/\tau_0 = \dot \gamma
\tau(c)$. The question is, to what extent the influence of
concentration on the polymer dynamics can be accounted for by a
concentration-dependent Weissenberg number. As we will see, this
concept applies well for all structural and dynamical properties.



\begin{figure}[t]
\begin{center}
\begin{tabular}{cc}
\begin{minipage}{1.65in}
\includegraphics*[width=\textwidth]
{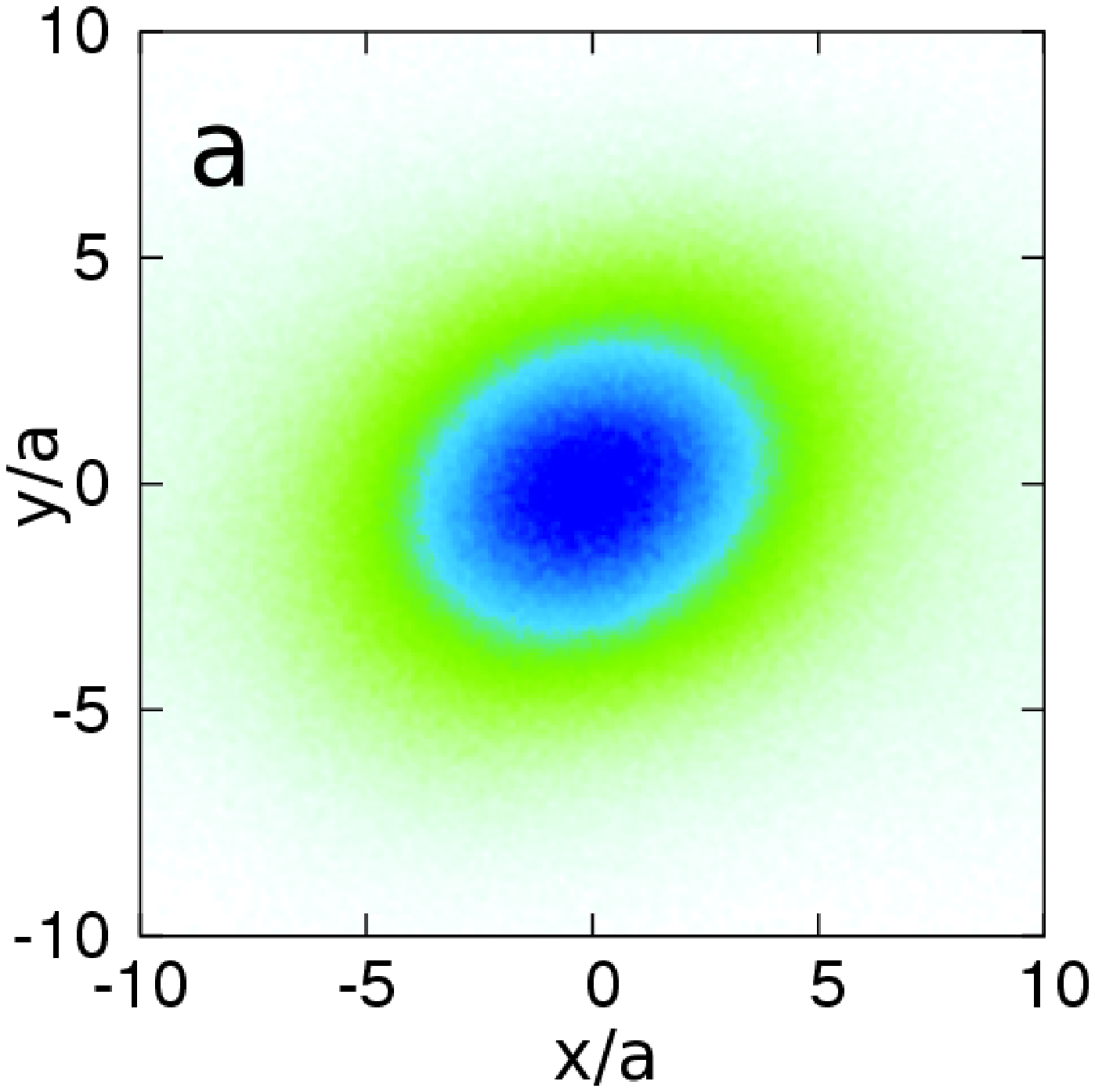}
\end{minipage}
&
\begin{minipage}{1.65in}
\includegraphics*[width=\textwidth]
{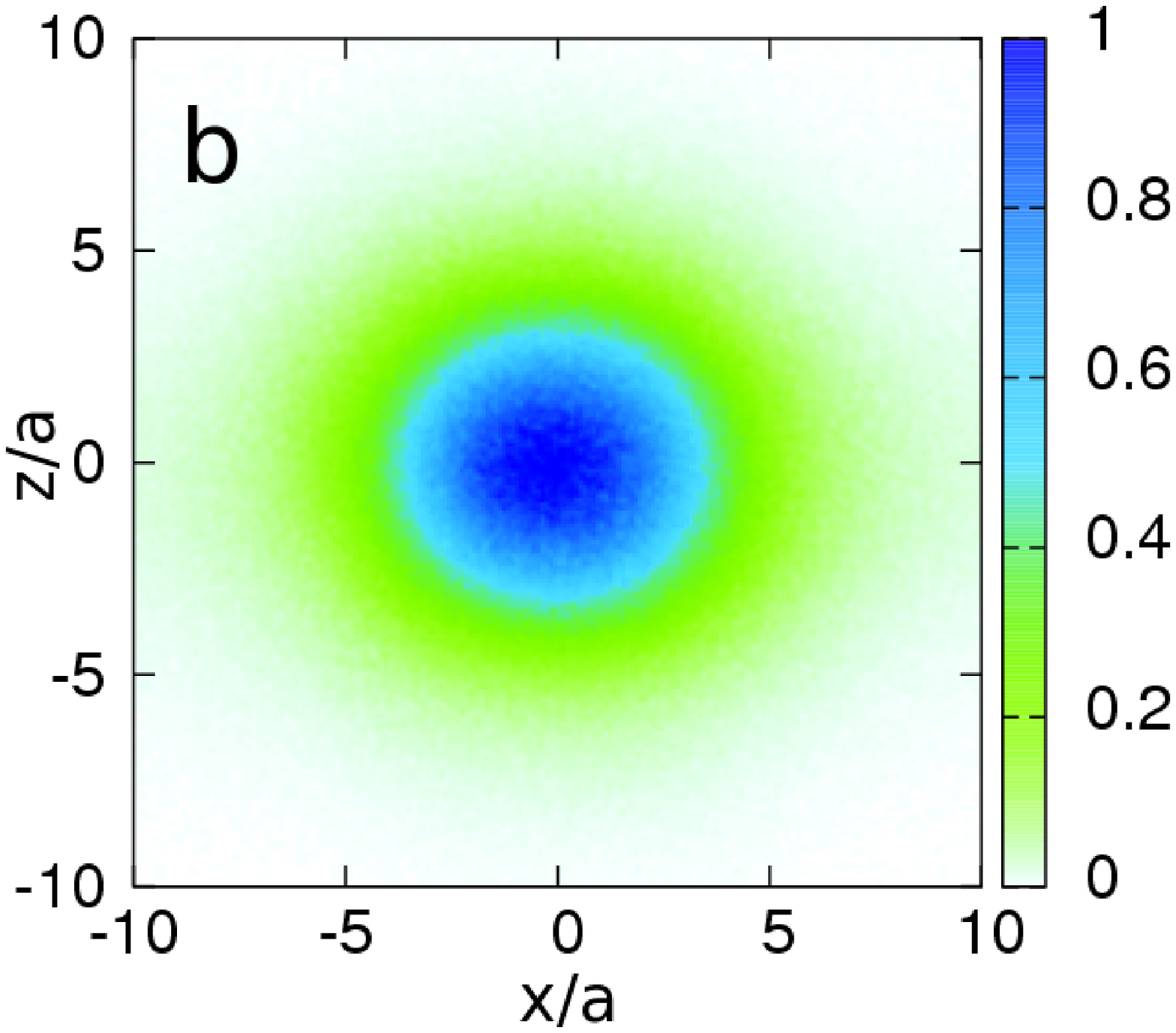}
\end{minipage}
\\
\begin{minipage}{1.65in}
\includegraphics*[width=\textwidth]
{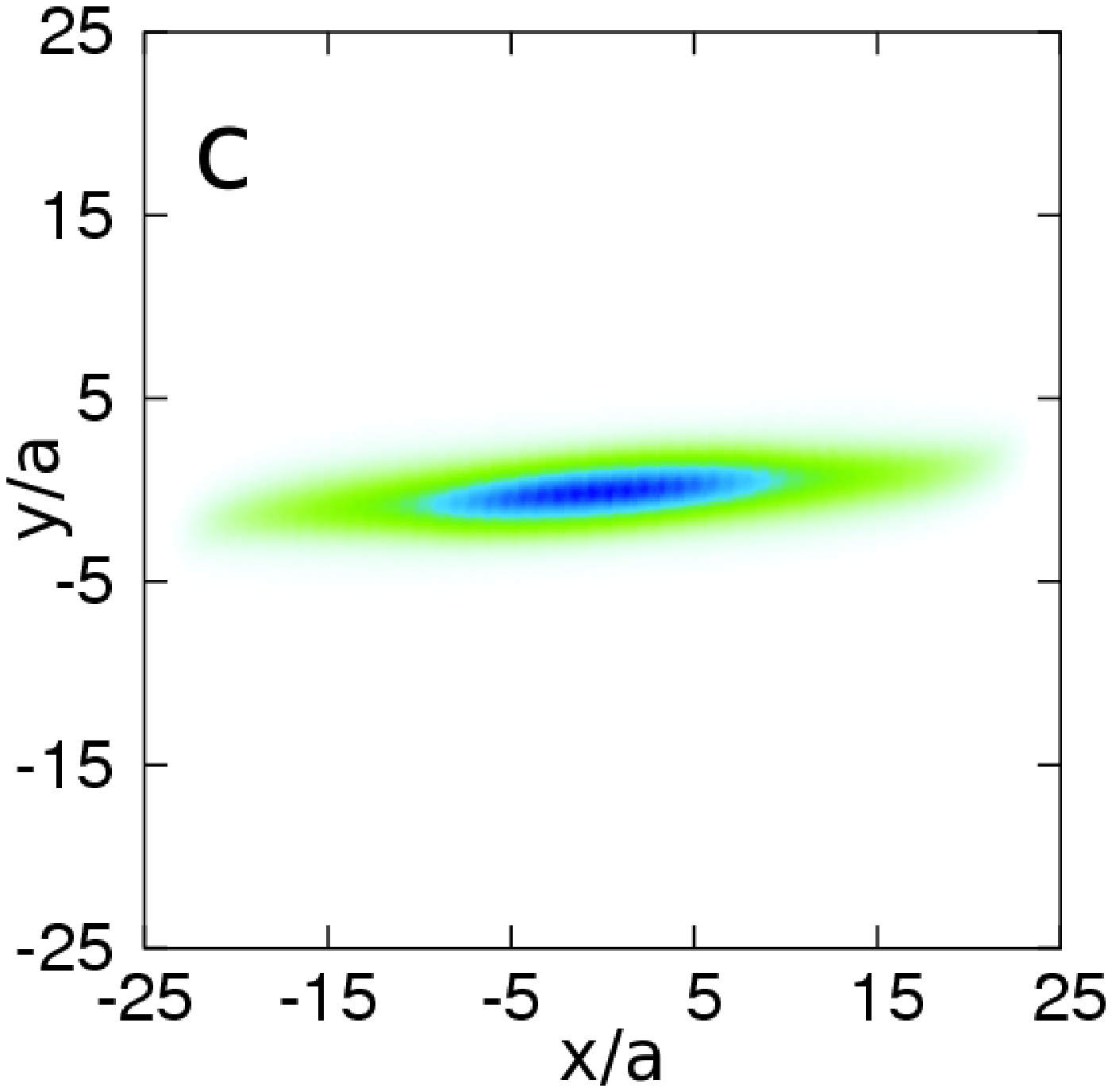}
\end{minipage}
&
\begin{minipage}{1.65in}
\includegraphics*[width=\textwidth]
{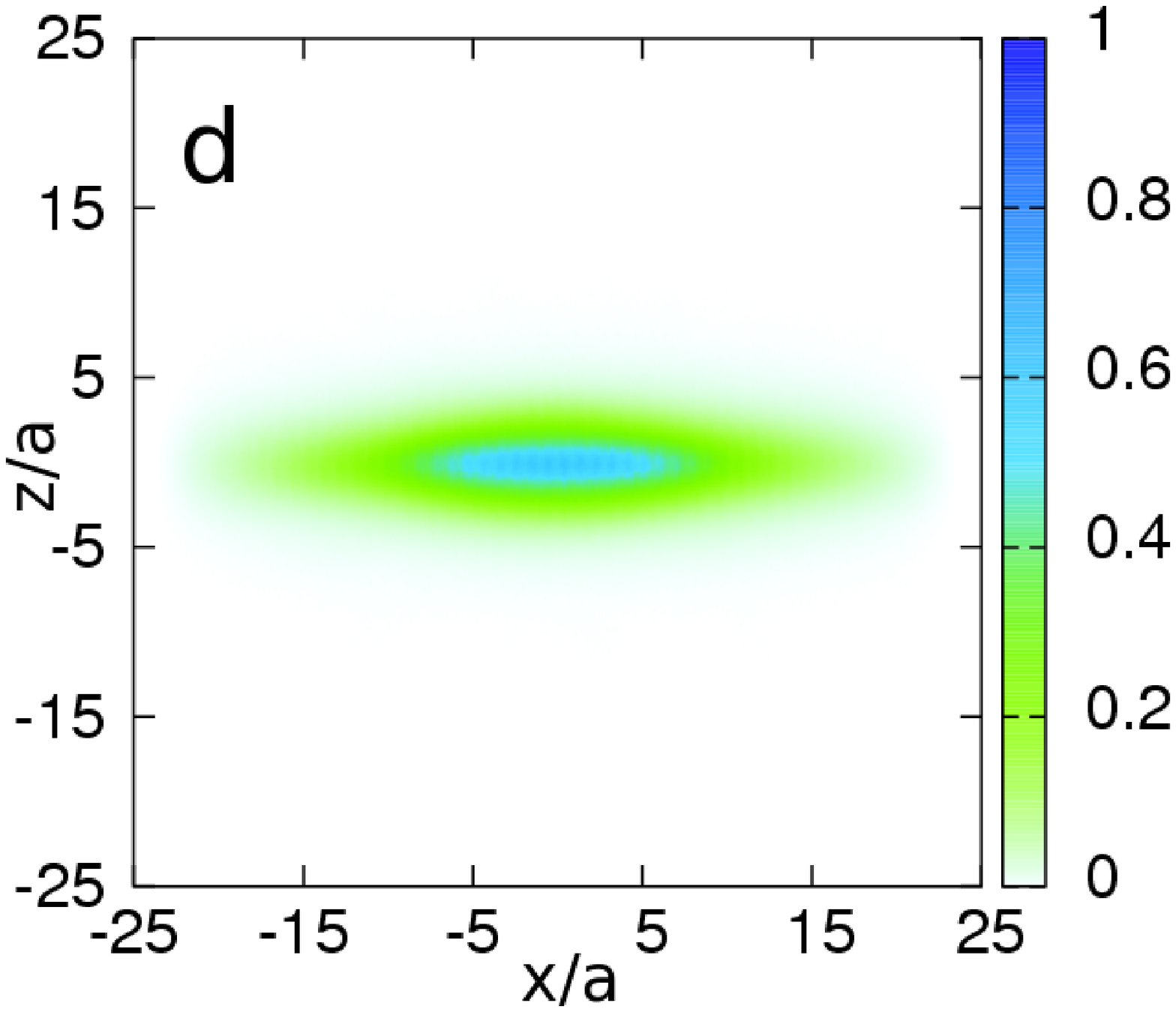}
\end{minipage}
\\
\end{tabular}
\caption{Monomer density distributions in the flow-gradient plane
(a), (c) and flow-vorticity plane (b), (d) for the Weissenberg
numbers $\mathrm{Wi}_c = 1$ (a), (b) and $\mathrm{Wi}_c = 307$
(c), (d). The concentration is $c/c^*=1.6$ and the chain length is
$N_m=50$. \label{shape_shear}}
\end{center}
\end{figure}
\begin{figure}[h]
\begin{center}
\begin{tabular}{cc}
\begin{minipage}{1.65in}
\includegraphics*[width=\textwidth]
{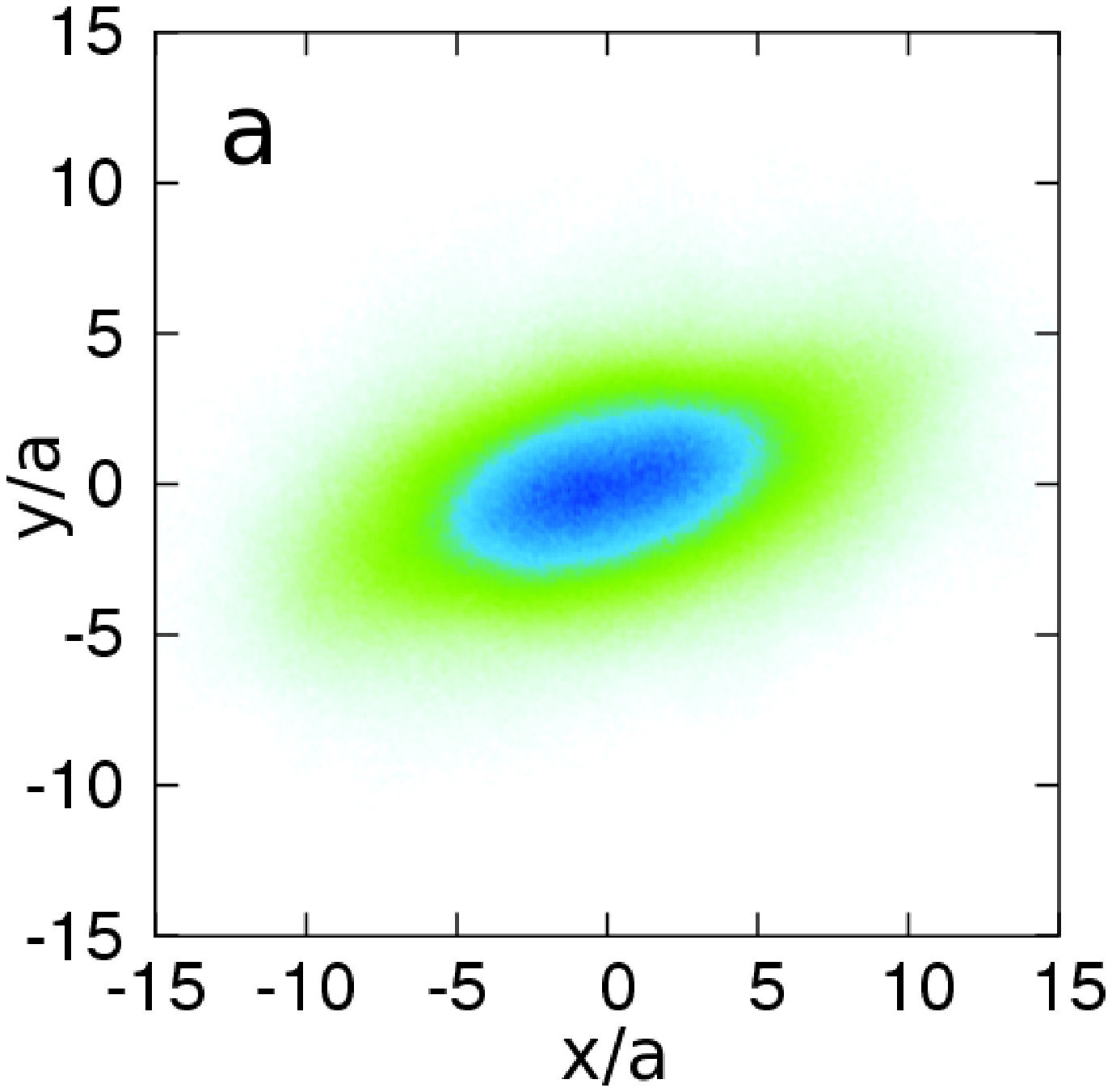}
\end{minipage}
&
\begin{minipage}{1.65in}
\includegraphics*[width=\textwidth]
{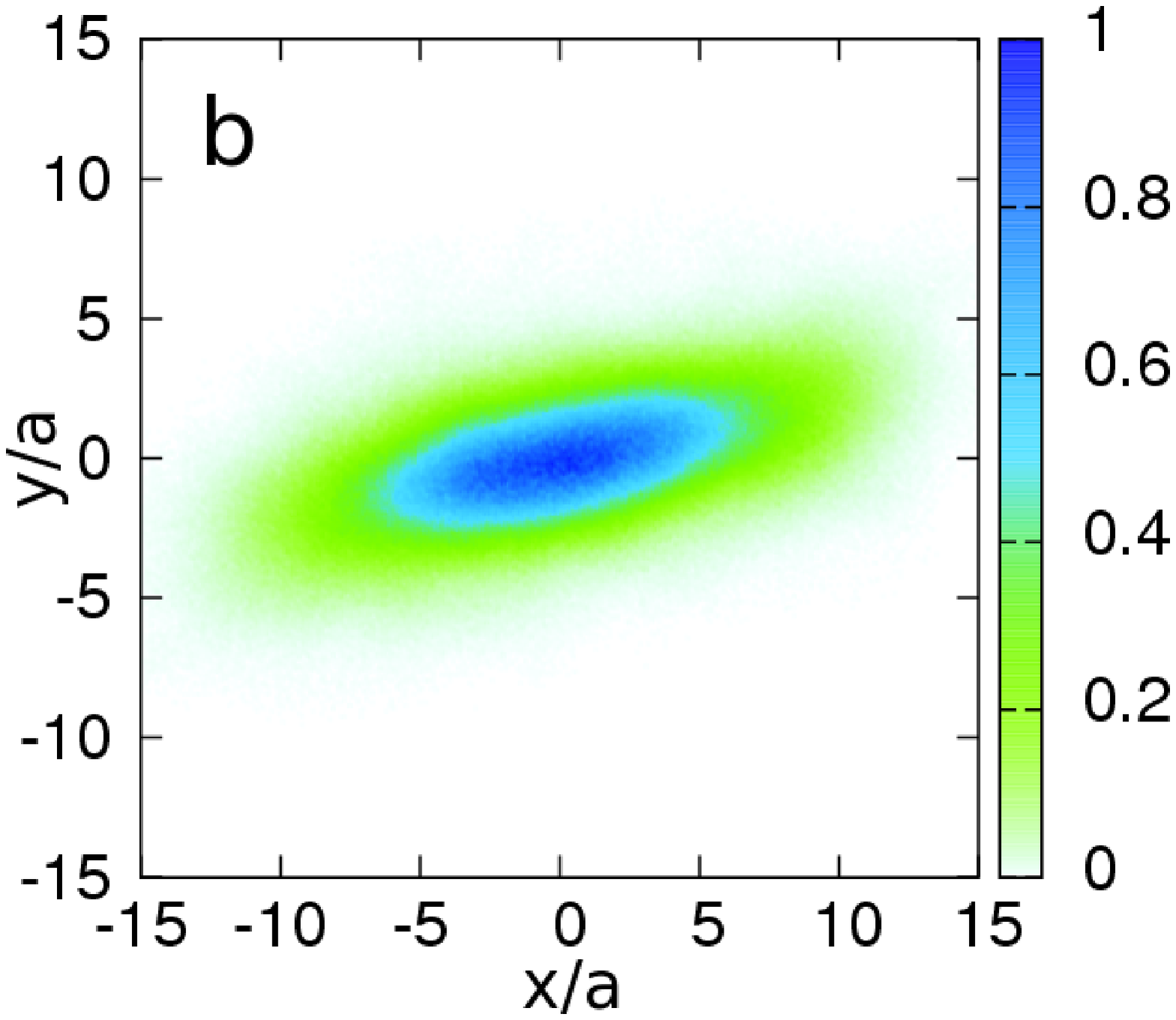}
\end{minipage}
\\
\end{tabular}
\caption{Monomer density distributions in the flow-gradient plane
for the shear rate $\tilde {\dot{\gamma}} = 10^{-3}$ and the
concentrations  $c/c^*=0.16$ (a) and $c/c^*=2.08$ (b),  which corresponds to the
Weissenberg numbers $\mathrm{Wi}_c = 6.2$  and $\mathrm{Wi}_c =11$, respectively.
The chain length is $N_m=50$.
\label{shape_c}}
\end{center}
\end{figure}

\begin{figure}[t]
\begin{center}
\includegraphics*[width=.4\textwidth,angle=0]{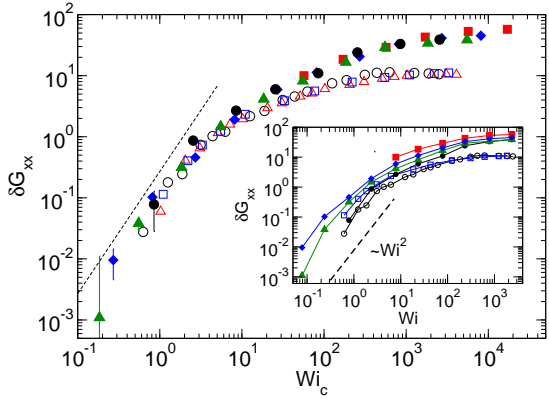}
\caption{Deformation ratios $\delta G_{xx}$ as function of
Weissenberg number. Open symbols correspond to
systems with $N_m=50$ for  $c/c^*=0.16$ ($\circ$),
$c/c^*=1.6$ (${\color{red}\triangle}$),  and $c/c^*=2.08$
(${\color{blue}\square}$). Filled  symbols denote results for
$N_m=250$ with the concentrations $c/c^*=0.17$ ($\bullet$), $c/c^*=2.77$
(${\color{green}\blacktriangle}$), $c/c^*=5.19$
(${\color{blue}\blacklozenge}$), and $c/c^*=10.38$
(${\color{red}\blacksquare}$). In the inset, the
same data are shown as function of $\wi$.
 \label{delta_G_xx}}
\end{center}
\end{figure}

\subsection{Conformations}

The average shape of an individual chain in solution under shear
is illustrated in Fig.~\ref{shape_shear} by the density
distribution of monomer positions with respect to the polymer
center of mass. At low Weissenberg numbers
(Fig.~\ref{shape_shear}(a), (b)), the polymers are only weakly
deformed and aligned with respect to the flow direction $x$,
whereas they are considerably stretched and aligned in the flow
direction and are compressed in the gradient and vorticity
direction for high shear rates (Fig.~\ref{shape_shear}(c), (d)).
Figure~\ref{shape_c} shows that the extent of deformation and
alignment depends upon polymer concentration. At the same
Weissenberg number, a larger deformation and a more pronounced
alignment is found at higher concentrations.

\subsubsection{Radius of gyration}

Polymer deformation and orientation are characterized
quantitatively by the gyration tensor
\begin{equation}
G_{\beta \beta'}=\frac{1}{N_m}\sum^{N_m}_{i=1}\langle
\Delta r_{i,\beta} \Delta r_{i,\beta'}\rangle,
\label{equ:Gtensor}
\end{equation}
where $\Delta r_{i,\beta}$  is the position of monomer $i$ in the
center-of-mass reference frame of the polymer.

In Fig.~\ref{delta_G_xx}, the relative deformation along the flow
direction
\begin{equation}
\delta G_{xx}=\frac{G_{xx}-G_{xx}^0}{G_{xx}^0},
\label{delta_G}
\end{equation}
where $G_{xx}^0=\langle R_g^2 \rangle/3$ is the gyration tensor at
equilibrium for the particular concentration, is shown for various
concentrations and polymer lengths. A significant polymer
stretching appears for $\wi_c
>1$. At large shear rates, the stretching
saturates at a maximum, which is smaller than the value
corresponding to a fully stretched chain ($G_{xx} \approx l^2
N_m^2/12$) and reflects the finite size of a polymer. This is
consistent with experiments on DNA \cite{schr:05,schr:05_1}, where
the maximum extension is on the order of half of the contour
length, and theoretical calculations \cite{wink:10}. It is caused
by the large conformational changes of polymers in shear flow,
which yields an average extension smaller than the contour length.
Nevertheless, molecules assume totally stretched conformations at
large Weissenberg numbers during their tumbling dynamics.
Interestingly, a universal dependence is obtained for $\delta
G_{xx}$ as function of a concentration-dependent Weissenberg
number $\wi_c$ at a given polymer length, whereas in terms of
$\wi$, polymers at larger concentrations exhibit a stronger
stretching at the same $\wi$, as shown in the inset of
Fig.~\ref{delta_G_xx} \cite{stol:05}. The latter is evident, since
the longest relaxation time of a polymer at higher concentrations
is larger and hence the polymer is more strongly deformed at the
same shear rate.

Theoretical calculations for single polymers in dilute solution
predict the dependence $\delta G_{xx} = C_{x}\wi^2$ for $\wi <1$,
where $C_x$ is a universal constant. The renormalization group
calculations of Ref. \cite{wang:90} yield $C_{x}=0.27$, whereas a
calculation based on a Gaussian phantom chain model yields $C_{x}
\approx 0.3$ \cite{brun:93,carl:94,pier:95,wink:06_1,wink:10}. As
shown in Fig.~\ref{delta_G_xx}, the simulations confirm the
quadratic dependence on the shear rate; $\delta G_{xx}$ is
independent of chain length for $\wi <1$ and $C_x \approx 0.1$.
For $\wi >10$, finite size effects appear and different asymptotic
values are assumed for the two chain lengths. We like to stress
that our simulations are in agreement with the molecular dynamics
simulation results of Ref.~\cite{pier:95} and the SANS data of
Refs.~\cite{lind:88,lind:89}.

\begin{figure}[t]
\begin{center}
\includegraphics*[width=.4\textwidth,angle=0]{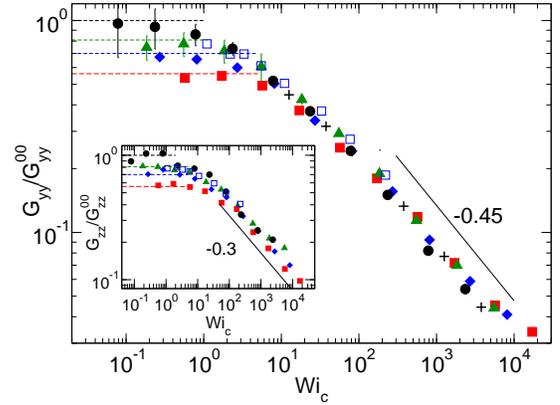}
\caption{
Relative deformations in the gradient and vorticity direction (inset).
Open symbols correspond to systems with $N_m=50$ for $c/c^*=2.08$
(${\color{blue}\square}$). Filled  symbols denote results for
$N_m=250$ with the concentrations $c/c^*=0.17$ ($\bullet$),
$c/c^*=1.38$ (+), $c/c^*=2.77$ (${\color{green}\blacktriangle}$),
$c/c^*=5.19$ (${\color{blue}\blacklozenge}$), and $c/c^*=10.38$
(${\color{red}\blacksquare}$). \label{delta_G_yy}}
\end{center}
\end{figure}

In the gradient and the vorticity directions, the polymers are
compressed, with a smaller compression in the vorticity direction
as shown in Fig.~\ref{delta_G_yy}. To highlight the universal
properties of the systems, we present the ratios $G_{\beta
\beta}/G_{\beta \beta}^{00}$ $(\beta \in \{y,z\})$, where
$G_{\beta \beta}^{00} = \left\langle R^2_{g0} \right\rangle/3$ is
calculated from the radius of gyration in dilute solution at
equilibrium.  At low shear rates---there is no detectable shear
deformation by shear---polymers shrink by concentration effects
for $c/c^* > 1$ (cf. Fig.~\ref{Rg_Rg0}). This is illustrated in
Fig.~\ref{delta_G_yy} for $\wi_c<10$ and various concentrations.
The dashed lines indicate the values of $\left\langle R^2_g
\right\rangle/\left\langle R^2_{g0} \right\rangle$ from
Fig.~\ref{Rg_Rg0}. Evidently, the ratios $G_{\beta
\beta}/G^{00}_{\beta \beta}$ are consistent with the values
$\left\langle R^2_g \right\rangle/\left\langle R^2_{g0}
\right\rangle$ for each concentration. The ratio for the shorter
chain $N_m=50$ and concentration $c/c^*=2.08$ is close to that of
the longer chain with a similar concentration ratio $c/c^*=2.77$.
This is consistent with the fact that  $\left\langle R^2_g
\right\rangle/\left\langle R^2_{g0} \right\rangle$ is independent
of chain length (cf. Fig.~\ref{Rg_Rg0}). With increasing shear
rate, the various curves progressively approach a universal
function, which decays as $\sim\mathrm{Wi}^{-0.45}$ over the
considered $\wi$-range. Hence, we obtain a different scaling
behavior of the radius of gyration along the flow direction and
the transverse directions. The reason is that a high monomer
density is maintained along the flow direction due to polymer
stretching, whereas the density in the transverse directions
decreases by flow-induced polymer shrinkage.

The exponents of the power-law decay of $G_{yy}$ and $G_{zz}$
compare well with the experimental data on single DNA
molecules~\cite{schr:05_1}. Similarly, simulations (with/without
hydrodynamic and excluded volume interactions) yield comparable
exponents~\cite{schr:05_1,hur:00}. However, simulations in Ref.
\cite{schr:05_1} for even larger Weissenberg numbers seem to
produce an exponent closer to the theoretically expected value of
$2/3$ \cite{wink:06_1,wink:10}. According to theory, there is a
broad crossover regime before the asymptotic behavior at large
Weissenberg numbers is assumed, and the considered $\wi$ fall into
that crossover regime.

\begin{figure}[t]
\begin{center}
\includegraphics*[width=.4\textwidth,angle=0]{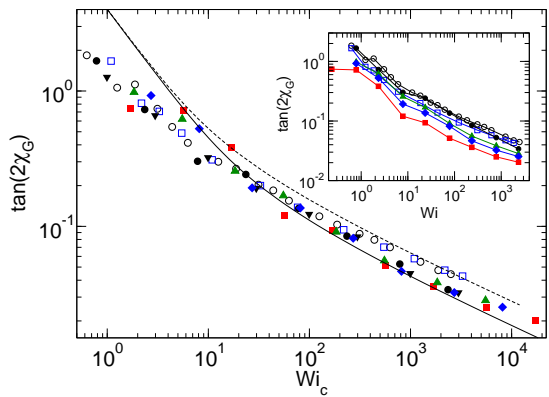}
\caption{Dependence of $\tan(2\chi_G)$ on shear rate.
Open symbols correspond to
systems with $N_m=50$ for  $c/c^*=0.16$ ($\circ$) and $c/c^*=2.08$
(${\color{blue}\square}$). Filled  symbols denote results for
$N_m=250$ with the concentrations $c/c^*=0.17$ ($\bullet$),
$c/c^*=0.35$ ($\blacktriangleleft$), $c/c^*=0.69$
($\blacktriangledown$), $c/c^*=2.77$
(${\color{green}\blacktriangle}$), $c/c^*=5.19$
(${\color{blue}\blacklozenge}$), and $c/c^*=10.38$
(${\color{red}\blacksquare}$). The solid and dashed lines are
theoretical results \cite{wink:06_1,wink:10}.
In the inset, the same data are shown as function of $\wi$. Lines are
guides for the eye only.}
\label{angle_G_1}
\end{center}
\end{figure}

\subsubsection{Alignment}

The alignment of the polymers is characterized by the angle
$\chi_G$, which is the angle between the eigenvector of the
gyration tensor with the largest eigenvalue and the flow
direction. It is obtained from the components of the radius of
gyration tensor via \cite{aust:99}

\begin{equation} \label{alignment}
 \tan(2\chi_G)=\frac{2G_{xy}}{G_{xx}-G_{yy}}.
\end{equation}

The dependence of $\tan(2\chi_G)$ on shear rate and concentration
is shown in Fig.~\ref{angle_G_1}. Again, a universal curve is
obtained for the different concentrations at a given polymer
length. Moreover, $\tan(2\chi_G)$ seems to be independent of
polymer length for $\mathrm{Wi}_c<100$, whereas we find a length
dependence  for larger Weissenberg numbers. In this high shear
rate regime, we find $\tan(2\chi_G) \sim (\mathrm{Wi}_c)^{-1/3}$.
We like to emphasize that only the shear rate can be scaled in
order to arrive at a universal function. The angle, or
$\tan(2\chi_G)$, cannot be scaled to absorb flow or polymer
properties in an effective variable. Hence, the universal behavior
of the alignment angle for various concentrations confirms that
the Weissenberg number $\wi_c$ is the correct scaling variable and
that the alignment of polymers at different concentrations depends
on the combination $\wi_c= \dot \gamma \tau$ of shear rate and
relaxation time only.

The analytical description of Refs.~\cite{wink:06_1,wink:10}
predicts the dependence
\begin{align} \label{align_theo}
\tan(2\chi_G)\sim \left(\frac{l_p}{L \wi^*}\right)^{1/3}
\end{align}
for semiflexible polymers in dilute solution in the limit $\wi^*
\to \infty$. Here, we introduce the Weissenberg number $\wi^* =
\dot \gamma \tau_{\mathrm{th}}$ for the theoretical result,
because the relaxation times from theory and simulation might not
be the same; $L$ is the length and $l_p$ the persistence length of
the polymer. The analytical result describes the simulation data
well at large shear rates, when the Weissenberg number of the
theoretical model is set to $\wi^*= \wi_c/2$. To compare the
predicted length dependence with that of the simulation, we apply
the relation $\langle R^2_{e0} \rangle = 2 l_pL$ to obtain a
persistence length, with $\langle R^2_{e0} \rangle$ the polymer
mean square end-to-end distance in dilute solution at equilibrium,
which yields $l_p/L \approx 0.025$ for $N_m=50$ and $l_p/L \approx
0.008$ for $N_m=250$. With these values, the ratio of
$\tan(2\chi_G)$ of the polymer of length $L=50a$ and $L=250a$ is
$1.5$. This compares well with the factor $1.33$ following from
the simulation results, which suggests that excluded volume
interactions are of minor importance for intermediate flow rates.

In the limit $\wi_c \to 0$, theory predicts $\tan(2\chi_G) \sim
\wi_c^{-1}$. The simulation data do not show this dependence on
the considered range of Weissenberg numbers, which might be due to
excluded volume interactions.

\begin{figure}[t]
\begin{center}
\includegraphics*[width=.4\textwidth,angle=0]{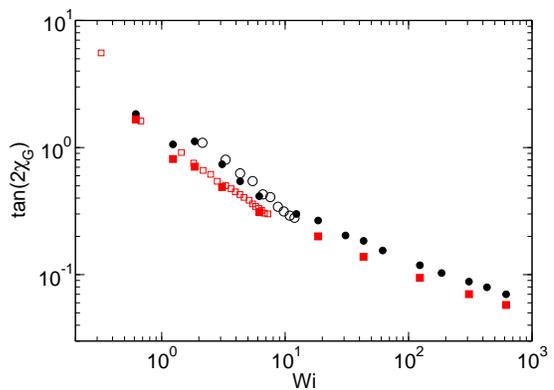}
\caption{Comparison of  experimental and simulation data
for the average orientation angle $\chi_G$. The concentrations are
$c=0.016/l^{-3}$ ($\bullet$),  $c=0.205/l^{-3}$
(${\color{red} \blacksquare }$)  for our simulations (with $N_m=50$),
and  $0.113g/l$ ($\circ$),  $1.094g/l$
(${\color{red}\square}$) for the experiments \cite{link:93}.}
\label{fig:cmp_chi_G}
\end{center}
\end{figure}

The inset of Fig.~\ref{angle_G_1} displays a strong dependence of
$\chi_G$ on concentration. The higher the concentration, the more
the chains are orientated along the flow direction. Such a
concentration effect has also been reported in light scattering
experiments~\cite{link:93}, where dilute polymer solutions are
considered. A comparison of the experimental data with the
simulation results is presented in Fig.~\ref{fig:cmp_chi_G}. Two
data sets are presented, a dilute solution, with the concentration
$0.113g/l$, and a semidilute solution with the approximately ten
times higher concentration $1.094g/l$, both taken from Fig.~8 of
Ref. \cite{link:93}. Evidently, the experimental data fit well
with our simulation results. Moreover, both, experiments and
simulations, yield a shift of the curves for the higher
concentrations to smaller Weissenberg numbers. In Ref.
\cite{link:93}, a factor $\beta_e \sim [\eta] \dot\gamma$, where
$[\eta]$ is the intrinsic viscosity, is used to present the data.
This quantity is proportional to $\wi$, since $[\eta]$ is
proportional to the longest relaxation time $\tau$; however,
$\beta_e$ and $\wi$ are not identical.  No adjustment parameter is
used in Fig.~\ref{fig:cmp_chi_G}, which suggests that the ratio
$\beta_e/\wi$ is close to unity.

\section{Rheology}

\subsection{Shear viscosity}

Under shear flow, the viscosity $\eta(\dot{\gamma})$ is obtained
from the relation
\begin{equation}
\eta(\dot{\gamma})=\frac{\sigma_{xy}(\dot{\gamma})}{\dot{\gamma}},
\label{equ:vis}
\end{equation}
where $\sigma_{xy}$ is the shear stress~\cite{bird:76,bird:87}. In
our simulations, $\sigma_{xy}$ is calculated using the virial
formulation of the stress tensor \cite{,alle:87,wink:09,wink:92}.
For sufficiently weak flow, the polymer solution is in the
Newtonian regime, i.e., $\sigma_{xy} \sim \dot{\gamma}$ and the
viscosity is independent of shear rate. Thus, the viscosity
obtained in this low shear rate regime is equal to the zero-shear
viscosity denoted by $\eta_0$. The latter depends on the monomer
concentration, which is often presented in the form
\begin{equation}
\eta_0=\eta_s[1+[\eta]c+k_H([\eta]c)^2+\ldots],
\label{equ:vis_0_d}
\end{equation}
where $\eta_s$ is the solvent viscosity and $k_H$ the Huggins
constant~\cite{bird:76,taka:85}.

We determine the intrinsic viscosity by a linear extrapolation to
zero concentration of both, $(\eta_0 -\eta_s)/ c \eta_s$ and the
inherent specific viscosity $[\ln(\eta_0/\eta_s)]/c$. The common
intercept of these two functions gives $[\eta]$
\cite{taka:85,pami:08}.

The intrinsic viscosity is proportional to $R_g^3/N_m$
\cite{doi:86} and is therefore proportional to the inverse of the
overlap concentration $c^*$~\cite{weil:79, rasp:95}. We find
$[\eta] c^* \approx 0.9$ and $[\eta] c^* \approx 1$ for the
polymer of length $N_m=40$ and $N_m=50$, respectively, which means
that the proportionality coefficient is close to unity for the
considered model systems.

The Einstein relation
\begin{align} \label{einstein}
\eta = \eta_s \left( 1 + 2.5 \phi \right) ,
\end{align}
where $\phi$ is the volume fraction, captures the concentration
dependence of hard sphere suspensions for $\phi \ll 1$. This
relation should also apply for dilute polymer solutions, when the
hydrodynamic radius $R_H$ is used to define the volume fraction,
i.e., $\phi=(4\pi/3) R_H^3 N_p/V$. Equations (\ref{equ:vis_0_d})
and (\ref{einstein}) are consistent if $[\eta] c^*= 2.5
(R_H/R_g)^3$. Since $[\eta] c^* \approx 1$, as explained above,
consistency requires $R_g/R_H \approx 1.36$. From our simulations,
we find the hydrodynamic radii $R_H \approx 3.8 l$ and $R_H
\approx 9.4 l$ for the polymers of length $N_m=50$ and $N_m=250$,
respectively, which yields the ratios $1.3$ and $1.36$. These
values are very close to the value necessary to match the Einstein
relation. The ratios are somewhat smaller than the asymptotic
value $R_g/R_H \approx 1.59$ for $N_m \to \infty$ obtained in
Ref.~[82], which is a consequence of the fact that we consider
insufficiently long chains.

\begin{figure}[t]
\begin{center}
\includegraphics*[width=.4\textwidth,angle=0]{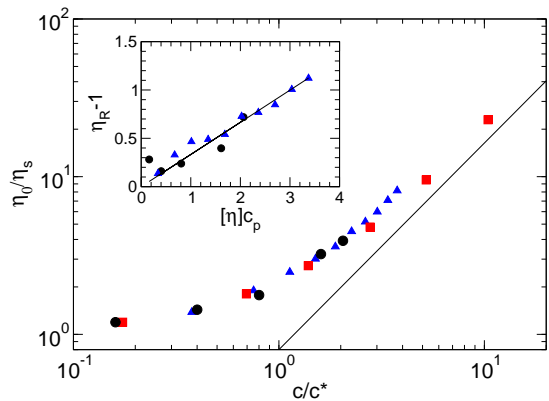}
\caption{Dependence of the zero shear viscosity on the scaled concentration
$c/c^*$ for the polymer lengths $N_m=40$ (${\color{blue} \blacktriangle }$),
$N_m=50$ ($\bullet$), and $N_m=250$ (${\color{red} \blacksquare }$).
The solid line indicates the power-law $(c/c^*)^{1/(3\nu-1)}$
with $\nu=0.6$. In the inset, $\eta_R-1$, Eq~(\ref{equ:vis_0_d2}), is shown
as function of  $[\eta]c$ for $N_m=40$ (${\color{blue} \blacktriangle
}$) and $N_m=50$ ($\bullet$); the slope of the solid line is $0.35$, which
corresponds to the Huggins constant of polymers in good solvent.
\label{fig:cmp_vis_0}}
\end{center}
\end{figure}

The term $k_H([\eta]c)^2$ depends on hydrodynamic interactions.
The value of the Huggins constant $k_H$ of flexible polymers is in
the range of $0.2-0.8$ and depends on solvent quality
\cite{taka:85,pami:08}. In good solvent, typically the value $0.3$
is found experimentally~\cite{pami:08}. Expressing
Eq.~(\ref{equ:vis_0_d}) in terms of the dimensionless parameter
$[\eta]c$ as
\begin{equation}
\eta_R = \frac{(\eta_0-\eta_s)}{\eta_s[\eta]c}=1+k_H[\eta]c+\ldots ,
\label{equ:vis_0_d2}
\end{equation}
which is denoted as relative viscosity, allows us to determine
$k_H$. The inset of Fig.~\ref{fig:cmp_vis_0} shows $\eta_R-1$ as
function of $[\eta]c$ for polymers of length $N_m=40$ and
$N_m=50$. The slope of the solid line is  $k_H=0.35$, in close
agreement with experiments \cite{taka:85,pami:08}.

For semidilute unentangled polymer solutions, the viscosity is
proportional to the number of blobs per chain and can be expressed
by the scaling relation~\cite{dege:79, taka:85, rasp:95, heo:05}
\begin{equation}
\eta_0=\eta_s \left( \frac{c}{c^*} \right)^{1/(3\nu-1)} .
\label{equ:vis_0_semi}
\end{equation}
Figure~\ref{fig:cmp_vis_0} displays zero-shear viscosities as
function of concentration for various polymer lengths. For
$c/c^*\gtrsim 3$, the data are close to the power-law of
Eq.~(\ref{equ:vis_0_semi}).

\begin{figure}
\begin{center}
\includegraphics*[width=.4\textwidth,angle=0]{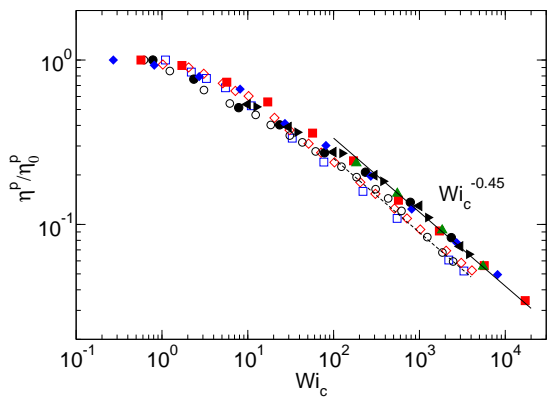}
\caption{Dependence of the polymer contribution to shear viscosity on shear rate.
Open symbols correspond to
systems with $N_m=50$ for  $c/c^*=0.16$ ($\circ$), $c/c^*=1.6$
(${\color{red}\Diamond}$), and $c/c^*=2.08$
(${\color{blue}\square}$). Filled  symbols denote results for
$N_m=250$ with the concentration $c/c^*=0.17$ ($\bullet$),
$c/c^*=0.35$ ($\blacktriangleleft$), $c/c^*=1.38$
($\blacktriangleright$), $c/c^*=2.77$
(${\color{green}\blacktriangle}$), $c/c^*=5.19$
(${\color{blue}\blacklozenge}$), and $c/c^*=10.38$
(${\color{red}\blacksquare}$).   } \label{vis_strong_shear}
\end{center}
\end{figure}

At sufficiently large shear rates, the polymers are aligned and
deformed, which implies shear
thinning~\cite{bird:87,aust:99,wink:06_1,wink:10}.
Figure~\ref{vis_strong_shear} shows the polymer contribution
$\eta^p$ to the shear viscosity. Similar to the alignment angle,
the viscosity is a universal function of the Weissenberg number
$\wi_c$ and shows a weak dependence on polymer length. It is
independent of shear rate for $\wi_c \ll 1$, decrease
approximately as $\wi_c^{-0.3}$ for $1 <\wi_c <10^2$, and
$\wi_c^{-0.45}$ for higher shear rates. This behavior is
consistent with other simulation results
\cite{schr:05_1,liu:89,pete:99, hsie:04,galu:10}. However, an even
stronger decay of the viscosity is observed in simulations at
larger shear rates in Refs. \cite{schr:05_1,aust:99}. Experiments
of dilute polymer solutions reported exponents ranging from $-0.4$
to $-0.85$~\cite{bird:87,schr:05_1}. Theoretical calculations for
dumbbells and finite extendable polymers predict the dependence
$\eta_p \sim \wi^{-2/3}$ in the limit $\wi \to \infty$
\cite{bird:87,oett:96,rubi:03,wink:06_1,wink:10}. The differences
in the observed behavior can be explained by a broad crossover
regime before the asymptotic behavior is reached.

The ratio of the viscosities of the two lengths is approximately
$1.33$ for the large Weissenberg-number regime, as for the
alignment angle, which compares well with the theoretically
predicted length dependence in Eq.~(\ref{align_theo}) (cf. Sec. IV
A. 2).

\subsection{Normal stress coefficient}

\begin{figure}[t]
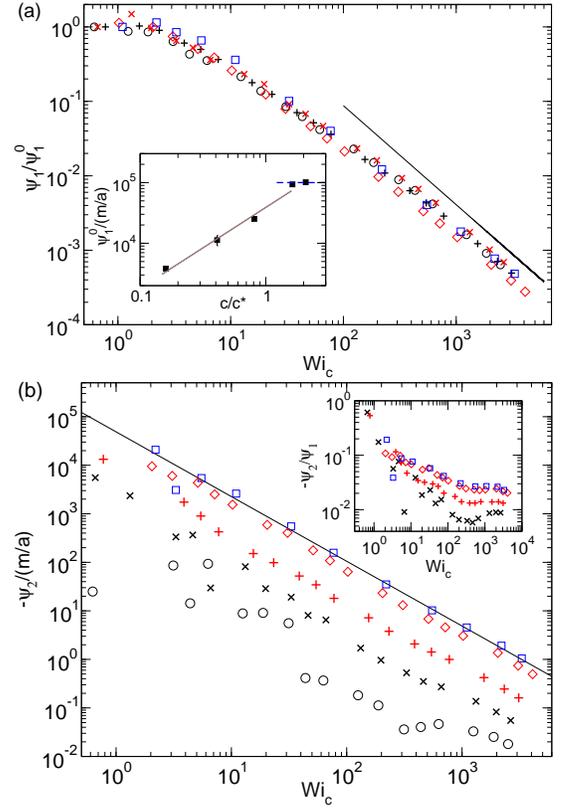

\includegraphics*[width=0.4\textwidth]{normal_str_1_2_rescal_1.eps}\\
\includegraphics*[width=0.4\textwidth]{normal_str_1_2_rescal_2.eps}
\caption{
First and second normal stress coefficients $\Psi_1$ (a) and $\Psi_2$ (b) for
polymers of length $N_m=50$ and
the concentrations $c/c^*=0.16$ ($\circ$), $c/c^*=0.41$
($\times$), $c/c^*=0.81$ (${\color{red}+}$), $c/c^*=1.63$
(${\color{red}\Diamond}$) and $c/c^*=2.08$
(${\color{blue}\square}$). The solid lines indicate the power-law decay
$\Psi_i \sim \wi_c^{-4/3}$. Inset in (a): Concentration
dependence of the zero-shear-rate first normal stress coefficient $\Psi_1^0$.
Inset in (b): Ratio of $\Psi_2/\Psi_1$ for the various concentrations.
\label{Psi}}
\end{figure}

The concentration and shear rate dependencies of the first and
second normal stress difference~\cite{bird:87,oett:96}
\begin{align} \label{stress_diff}
\Psi_1 = & (\sigma_{xx}-\sigma_{yy})/\dot{\gamma}^2 , \\
\Psi_2 = & (\sigma_{yy}-\sigma_{zz})/\dot{\gamma}^2
\end{align}
are displayed in Fig.~\ref{Psi}.  Within the accuracy of the
simulations, the ratio $\Psi_1/\Psi_1^0$, where $\Psi_1^0$ is the
stress difference at zero shear rate, is an universal function of
$\wi_c$ for various concentrations and decreases as $\Psi_1 \sim
\dot{\gamma}^{-4/3}$ for large shear rates. This is consistent
with analytical calculations~\cite{bird:87,oett:96,wink:10},
various computer
simulations~\cite{zylk:91,oett:96,schr:05_1,liu:89,pete:99,
hsie:04}, and experiments~\cite{teix:05, schr:05} for dilute
solutions. Similar to the viscosity, the decrease is related to
the finite polymer extensibility. Both, hydrodynamic and excluded
volume interactions contribute to $\Psi_1$, which is shown in Ref.
\cite{pete:99} for single polymers. Here, we find the power law
\begin{align}
\Psi_1^0 \sim \left( \frac{c}{c^*} \right)^{1.3},
\end{align}
as shown in the inset of Fig.~\ref{Psi}(a). Hence, the first
normal stress difference exhibits a significant dependence on
excluded volume interactions. At large concentrations, $\Psi_1^0$
might saturate; at least, we cannot exclude such a saturation at
the upper limit of the considered concentration range.

Second normal stress differences are presented in
Fig.~\ref{Psi}(b) for various concentrations. At low
concentrations, their values are much smaller than those of
$\Psi_1$, and hence cannot be calculated within the same accuracy,
and the values $\Psi_2^0$ are difficult to find. We therefore
present the simulations results for $\Psi_2$ directly rather than
in scaled form. Similar to $\Psi_1$, the second normal stress
difference decreases as $\Psi_2\sim\dot{\gamma}^{-4/3}$ with
increasing shear rate. Again, excluded volume and hydrodynamic
interactions contribute to $\Psi_2$
\cite{bird:87,oett:96,pete:99}. The ratio $\Psi_2/\Psi_1$ is
concentration dependent, as shown in Fig.~\ref{Psi}(b). At small
$\wi_c$ and large concentrations, the ratio is close to unity,
decreases with increasing shear rate and assumes a constant value
above a certain $\wi_c$, which seems to depend on concentration.
The plateau value itself increases with increasing concentration.
A similar plateau has been found in simulations of dilute
solutions in Ref.~\cite{hsie:04}. The concentration dependence of
the plateau value suggests that excluded volume interactions
determine the behavior of the normal stress differences. If
hydrodynamic interactions would be dominant, we would expect a
decrease of the plateau with increasing concentration due to
screening of hydrodynamic interactions by polymer overlap.

\section{Conclusions}

We have calculated conformational, dynamical, and rheological
properties of polymers in dilute and semidilute solution under
shear flow by mesoscale hydrodynamic simulations. At equilibrium,
our simulations confirm the scaling predictions for the
concentration dependence of the radius of gyration and the longest
relaxation time. Moreover, we find signatures for the screening of
excluded volume interactions in the static structure factor.

In shear flow, the polymers exhibit deformation---the polymer is
stretched in flow direction and shrinks in the transverse
directions---and alignment, which depend on shear rate and
concentration. As one of the the main results of the paper, we
have shown that the relative deformation $\delta G_{xx}$ in the
flow direction, the alignment $\tan(2 \chi_G)$, and the viscosity
$\eta/\eta_0$ are universal functions of the
concentration-dependent Weissenberg number $\wi_c = \dot \gamma
\tau(c)$ \cite{galu:10,hur:01}. This is surprising because $\tau$
increases rapidly with increasing concentration. Moreover, it
indicates that the dynamics under shear flow is still governed by
the relaxation time at equilibrium despite the anisotropic
deformation of a polymer. This is not evident a priori, as
expressed by the scaling behavior of the radius of gyration tensor
components $G_{yy}$ and $G_{zz}$. Here, we find a
concentration-independent scaling behavior at large $\wi_c$ only
when these values are scaled by their equilibrium values in {\em
dilute solution}. Hence, the deformations transverse to the flow
directions seem to exhibit a scaling behavior corresponding to a
dilute solution, however, with the relaxation time of the
concentrated system.

In addition, the zero-shear viscosity obeys the scaling
predictions with respect to the concentration dependence.
Moreover, for the first time, we show by simulations that the
Huggins constant is equal to  $k_H=0.35$ for a flexible polymer in
good solvent, which is in close agreement with experimental
results \cite{pami:08}.

Finally, we find a strong concentration dependence of the normal
stress differences. Their ratio shows that intermolecular excluded
volume interactions determine their behavior at all shear rates.

Our simulations reveal a complex interplay between shear rate,
deformation, and intramolecular excluded volume interactions,
which is difficult to grasp by analytical theory.

\begin{acknowledgments}
The financial support by the Deutsche Forschungsgemeinschaft (DFG)
within SFB TR6 (project A4) is gratefully acknowledged. We are
grateful to the J\"ulich Supercomputer Centre (JSC) for allocation
of a special CPU-time grant.
\end{acknowledgments}


\newpage



\end{document}